\documentclass[twocolumn,showpacs,preprintnumbers,amsmath,amssymb]{revtex4}

\usepackage{graphicx}% Include figure files
\usepackage{bm}% bold math

\begin{document}
\draft

%\preprint{PRE}

\title{Creep motions of flux lines in type II superconductors with point-like defects}

\author{Meng-Bo Luo\(^{\dagger,\ddagger}\), Xiao Hu\(^{\dagger,\ast}\)
and Valerii Vinokur\(^{\star}\) }

\affiliation{ \(^{\dagger}\)WPI Center for Materials
Nanoarchitectonics, National Institute for Materials Science,
Tsukuba 305-0047, Japan
\\ \(^{\ddagger}\)Department of Physics, Zhejiang University, Hangzhou 310027,
China
\\ \(^{\star}\)Materials Science Division, Argonne National
Laboratory, Illinois 60439, USA}

\date{\today}

\begin{abstract}
We simulated the creep motions of flux lines subject to randomly
distributed point-like pinning centers. It is found that at low
temperatures, the pinning barrier $U$ defined in the Arrhenius-type
$v-F$ characteristics increases with decreasing force $U(F) \propto
F^{-\mu}$, as predicted by previous theories. The exponent $\mu$ is
evaluated as $0.28\pm 0.02 $ for the vortex glass and $\mu\simeq
0.5\pm 0.02$ for the Bragg glass (BrG). The latter is in good
agreement with the prediction by the scaling theory and the
functional-renormalization-group theory on creep, while the former
is a new estimate. Within BrG, we find that the pinning barrier is
suppressed when temperature is lifted to approximately half of the
melting temperature. Characterizations of this new transition at
equilibrium are also presented, indicative of a phase transition
associated with the
replica-symmetry breaking. \\
 \\
\end{abstract}

%\pacs{74.25.Qt, 64.60.Ht, 05.70.Ln}% PACS, the Physics and Astronomy
                             % Classification Scheme.
%\keywords{} %Use showkeys class option if keyword
                             %display desired
\maketitle

\noindent {\it Introduction --} An elastic medium immersed into
random environment is a generic system modeling a wealth of physical
situations. In spite of the considerable efforts expended and
impressive progress, many crucial issues of the underlying physics
remain unresolved, and understanding the statics and dynamics of
such systems is one of the major challenges of condensed matter
physics.

Intensive activities have taken place in vortex states of type II
superconductors since the discovery of high-$T_c$ superconductivity
in cuprates
\cite{Blatter_RMP,Ioffe1987,Vinokur2000,FRG,Review_Nattermann2000,Review_Giamarchi,Review_Nattermann2004}.
Initiated by the collective pinning theory \cite{LO}, theoretical
understanding has been advanced
\cite{Ioffe1987,Nattermann1987,Fisher1989,Feigelman1989,BrG_Nattermann,FRG}.
The competition between the elastic force and the random force
builds a complex potential landscape governing vortex dynamics. An
important notion has been established, namely the typical energy
barrier felt by the system becomes divergingly large $U(F) \sim
F^{-\mu}$ at small driving forces
\cite{Ioffe1987,FRG,Feigelman1989,BrG_Nattermann}, which results in
an extremely small velocity of the system following the Arrhenius
law.

The degree of divergence $\mu$ is governed by the large-scale
elastic properties of the system and properties of disorder
\cite{FRG}. The value of exponent $\mu$ is however not easy to
evaluate accurately, since the static and dynamic properties of the
system are intervened in a very complex way. The situation is even
more severe when the randomness is large and thus the order is
destroyed fully. In most theory restoring forces assume elastic
behaviors of the system involved, which is not the case in strongly
disordered systems, where plastic deformations become important (See
discussions in Ref.\cite{Vinokur2000}).

We tackle this problem by computer simulations on three-dimensional
(3D) flux lines subject to randomly distributed point-like pinning
centers (for a parallel work on 1D domain wall in 2D space see
Refs.~\cite{Lemerle_1998,Kolton}). Tuning the strength of random
pinning force, we can reach the Bragg glass (BrG)
\cite{BrG_Nattermann,BrG_GD} at weak pinning and vortex glass (VG)
\cite{VG} at relatively strong pinning, at equilibrium. Langevin
dynamics then permits us to explore the dynamics of the system
subject to driving force at various temperatures without any
uncontrollable approximation.

The main results are as follows: The creep law of flux lines
predicted by previous theories is clearly reproduced. The exponent
is estimated as $\mu=0.5\pm 0.02$ for BrG and $\mu=0.28\pm 0.02$ for
VG, both universal for the respective class. While the former one is
very close to the expected value, the latter one is a new
estimation. For weak pinning where BrG is stable below the melting
temperature $T_m$, we find an unpinned state at $T_g<T<T_m$ in
addition to the pinned one at $T<T_g$, indicating a replica-symmetry
breaking transition at $T_g\simeq T_m/2$.

\vspace{3mm}

 \noindent {\it Model and simulation details --} The model system is a stack
of superconducting planes of thickness $d$ with period $s$ of the
layer structure, with the magnetic field perpendicular to the
layers. Each plane contains $N_v$ vortices and $N_p$ quenched pins.
The overdamped equation of motion of the \(i\)th vortex at position
${\bf r}_i$ is \cite{model,PRL2007_LH}

\begin{equation}
   \eta \dot{{\bf r}}_i = -\sum\limits_{j\neq i} \nabla_i
   U^{VV}({\bf r}_{ij})- \sum\limits_{p} \nabla_i
   U^{VP}({\bf r}_{ip}) + {\bf F} + {\bf F}_{th} .
\end{equation}
Here $\eta$ is the viscosity coefficient. The intraplane vortex
repulsion is given by the modified Bessel function
$U^{VV}(\rho_{ij},z_{ij}=0) = d\epsilon_0
K_0(\rho_{ij}/\lambda_{ab})$, and the interplane vortex attraction
is $U^{VV}(\rho_{ij},z_{ij}=s)=(s\epsilon_0 /\pi) [1+\ln
(\lambda_{ab}/s)][(\rho_{ij}/2r_g)^2-1]$ for $\rho_{ij} \leq 2r_g$
and $U^{VV}(\rho_{ij},z_{ij}=s)=(s\epsilon_0 /\pi) [1+\ln
(\lambda_{ab}/s)][\rho_{ij}/r_g-2]$ otherwise, between two vortices
belonging to the same flux line and sitting on adjacent planes,
where $\epsilon_0=\phi_0^2/2\pi\mu_0\lambda_{ab}^2$ with
$\lambda_{ab}$ the magnetic penetration depth of the superconducting
layer, $r_g = \gamma s$ with $\gamma$ the anisotropy parameter. The
pinning potential is $U^{VP}(\rho_{ip}) = -\alpha
A_p\exp[-(\rho_{ip}/R_p)^2]$, where $A_p = (\epsilon_0 d/4)\ln [1 +
(R_p^2/2\xi_{ab}^2)]$ with \(\xi_{ab}\) the in-plane coherence
length and $\alpha$ the dimensionless pinning strength. Finally,
${\bf F}$ is the uniform Lorentz force, and ${\bf F}_{th}$ is the
thermal noise force with zero mean and a correlator $\langle
F^p_{th}(z,t) F^q_{th}(z^\prime,t^\prime)\rangle =2\eta
T\delta^{pq}\delta (z-z^{\prime })\delta (t-t^{\prime })$ with $p,
q= x, y$. The units of time is $\tau_0=\eta
\lambda_{ab}^2/d\epsilon_0\simeq 0.03 ns$ for $\lambda_{ab} = 2000
\AA$ and $\eta /d \simeq 10^{-8} Pa\cdot s $ typically for BSCCO
\cite{Bulaevskii_2004}. Conventions including the units of physical
quantities are taken same as Ref.~\cite{PRL2007_LH}.

\vspace{3mm}

\noindent {\it Creep motions --} We start with the case of
$\alpha=0.2$, for which an amorphous VG is realized at low
temperature \cite{PRL2007_LH}. Figure 1 presents the average
velocity at small driving forces at several typical temperatures.
The driving forces are limited to $F < F_{c0}/2$ with $F_{c0} =
0.232$ the zero-temperature depinning force \cite{PRL2007_LH}, where
the creep feature of vortex motion is profound. At even lower
temperature, the dynamics is to be governed by the fixed point at
$F_{c0}$, and thus omitted in Fig.~1 (see
Refs.~\cite{Blatter2004,PRL2007_LH}). The velocity $v = 0.01$
corresponds approximately to $60m/s$.

\begin{figure}[tu1]
\begin{center}
\includegraphics[width=7.5cm]{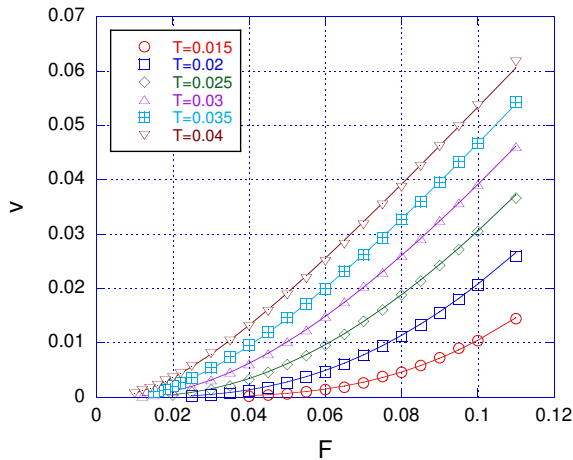}
\caption{Velocity-force ($v-F$) characteristics at different
temperatures for $\alpha=0.2$. The solid lines are fitting curves
based on Eq.~(\ref{eqn:creeplaw}) as described in text. The system
size is $L_x \times L_y \times N_z =30\times 30\times 20$ with the
lateral directions measured by $\lambda_{ab}$. Errors and
finite-size effects are smaller than the size of symbols.}
\end{center}
\end{figure}

The $v-F$ characteristics in Fig.~1 are described well by the
Arrhenius-type functions at respective temperatures

\begin{equation}
v=v_0 \exp \left [ -\frac{U}{T} \left( \frac{F_{c0}}{F}
\right)^{\mu} \right
   ].
\label{eqn:creeplaw}
\end{equation}

\noindent From the least-squares fitting, we find $U = 0.15 \pm
0.02$ and $\mu = 0.28 \pm 0.02$, with $F_{c0}=0.232$ given in the
previous work \cite{PRL2007_LH}. The prefactor $v_0$ depends on
temperature roughly in a power-law way. For a stronger pinning
$\alpha = 0.4$, it is found $U = 0.30 \pm 0.03$ and $\mu = 0.28 \pm
0.02$ (raw data not shown here). Therefore, the exponent $\mu$ is
universal for VG.

We then turn to the weaker pinning force $\alpha = 0.1$ where BrG is
stable at low temperatures $T\le T_m\simeq 0.075$. The $v-F$
characteristics are displayed in Fig. 2. Carrying on the same
analysis mentioned above, we obtain $U = 0.034 \pm 0.006$ and $\mu =
0.50 \pm 0.02$. For $\alpha = 0.05$, for which the equilibrium state
is also a BrG at low temperatures, we obtain $U = 0.018 \pm 0.003$
and $\mu = 0.50 \pm 0.05$. The above results indicate that the
exponent $\mu$ for $\alpha\le 0.1$ falls into another class,
universal for weak pinning strengths for which BrG is realized at
low temperatures.

\begin{figure}[tu2]
\begin{center}
\includegraphics[width=7.5cm]{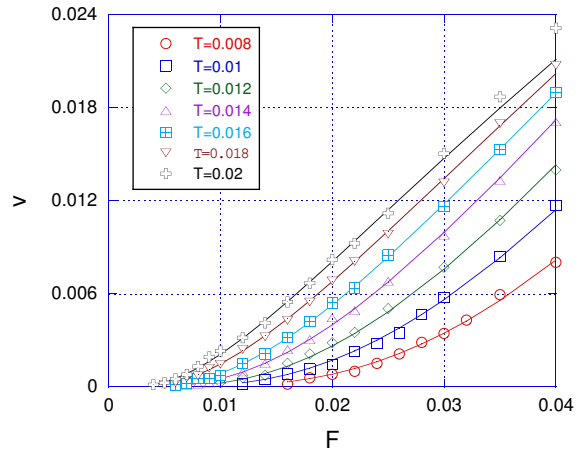}
\vspace{2mm}\caption{$v-F$ characteristics same as Fig.~1 except for
$\alpha=0.1$.}
\end{center}
\end{figure}

 Our estimate on the exponent $\mu$ is in
good agreement with the theoretical prediction
\cite{Feigelman1989,FRG}
\begin{equation}
 \mu = \frac{D-2+2\zeta_{\rm eq}}{2-\zeta_{\rm eq}}=\frac{1}{2}
\label{eqn:valuemu}
\end{equation}
with $\zeta_{\rm eq} = 0$ the roughness exponent for BrG.

 As pointed out in Ref.~\cite{Vinokur2000}, previous theories on
vortex creep started from the hypothesis of elastic energy, which is
established only for elastic deformations, and thus work at best for
BrG. The exponent $\mu$ should be reformulated for strong pinnings
associated with VG where deformations are plastic
\cite{Vinokur2000}. It was not clear yet that whether $\mu$ is
larger in BrG or in VG \cite{PRB1997_GD,Vinokur2000}. The present
simulation results suggest a stronger divergence of energy barrier
in BrG.

\vspace{3mm}

\noindent {\it New phase boundary in BrG --} So far, we have
concentrated on low temperatures, where the thermal energy only
activates flux lines from pinning centers, resulting in an
Arrhenius-type motion with a well defined energy barrier which
depends on the driving force. As temperature is lifted, thermal
fluctuations become more important, which makes the competition
between the randomness and the intervortex forces very subtle. Here
we focus on the weak pinning case $\alpha=0.05$, for which the
ground state is a BrG with $T_m\simeq 0.075$.

\begin{figure}[tu3]
\begin{center}
\includegraphics[width=7.5cm]{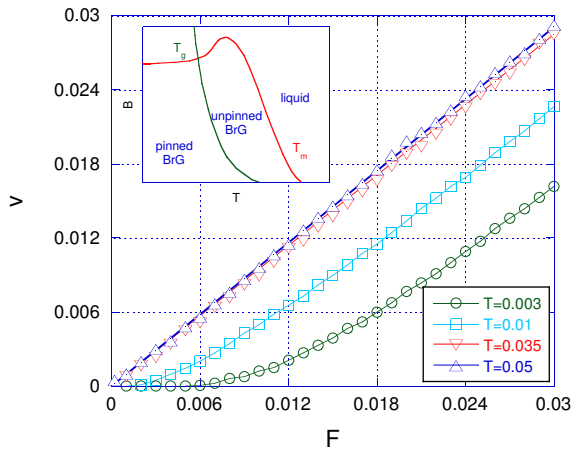}
\caption{$v-F$ characteristics for $\alpha= 0.05$ at different
temperatures. A transition temperature $T_g\simeq 0.035$ is
defined which separates the nonlinear and linear $v-F$ curves. The
solid lines are for eye-guide. Inset: a schematic phase diagram of
vortex states.}
\end{center}
\end{figure}

Figure 3 presents the $v-F$ characteristics at several typical
temperatures. At low temperatures, the $v-F$ curves are nonlinear at
low forces, similar to Figs. 1 and 2. However, above $T_g\simeq
0.035$, the $v-F$ characteristics is linear down to the small force
limit. This indicates that the potential barrier sensed by BrG is
smeared to zero in an intermediate temperature regime $T_g<T<T_m$,
where the crystalline order is still preserved. A similar change is
observed for relative strong pinning where VG is realized at low
temperatures, for which the details will be reported elsewhere.

\begin{figure}[tu4]
\begin{center}
\includegraphics[width=8.5cm]{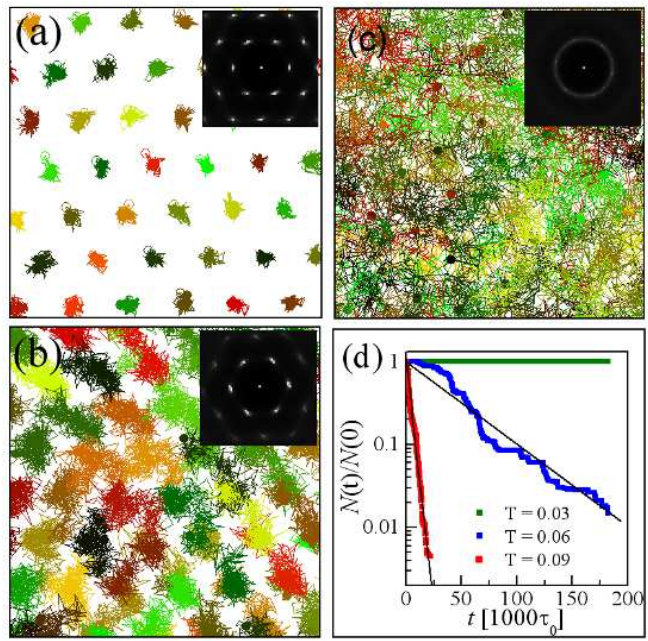}
\caption{Trajectories of vortices on a typical layer and structure
factors averaged on all layers at three typical temperatures: (a):
$T = 0.03$, (b): $T=0.06$, and (c): $T=0.09$. Only a part
($L_x\times L_y=14 \times 14$) of the totally simulated system is
displayed. The time interval is $100\tau_0$ and the total time shown
here is $50,000\tau_0$. (d): temporal evolution of the number of
vortices which are trapped in the initial cages.}
\end{center}
\end{figure}

We have found that even at equilibrium, i.e. for zero driving force,
the system behaves in qualitatively different ways for temperatures
below and above $T_g$. In Fig.~4, we display two-dimensional vortex
trajectories and the corresponding structure factors. At $T = 0.03(<
T_g)$ (Fig.~4(a)), vortices are trapped in cages formed by the
random pinning potential and the intervortex repulsions (even for
time much longer than that shown in Fig.~4); a profound BrG order is
clearly seen. At $T = 0.09(> T_m)$ (Fig.~4(c)), the system is in a
liquid state where vortices diffuse freely and randomly. At
$T=0.06$, the behaviors of the vortices are different as evidenced
in Fig.~4(b): Over an intermediate time scale, vortices are trapped
at local positions in a way such that the BrG order is established.
Vortices then move quickly to another set of localized positions,
which also establishes the BrG order, and stay in the new positions
over another intermediate time duration. The intermittent
trapped-and-moving motions continue, and eventually vortices can
travel over distances larger than the vortex lattice constant $a_0$.

In order to capture the vortex motions better, we have monitored the
number $N(t)$ of vortices that move over distances smaller than
$a_0$ after time $t$. As shown in Fig.~4(d), $N(t)$ decreases
exponentially with $t$ in a large time scale at $T=0.06$ since
vortices move over distances larger than the lattice constant. The
exponential decay occurs even in a small time scale at $T=0.09$ as
the system is random. At $T=0.03$, $N(t)$ remains unity during the
time evolution because all the vortices are caged.

\begin{figure}[b]
\begin{center}
\includegraphics[width=7cm]{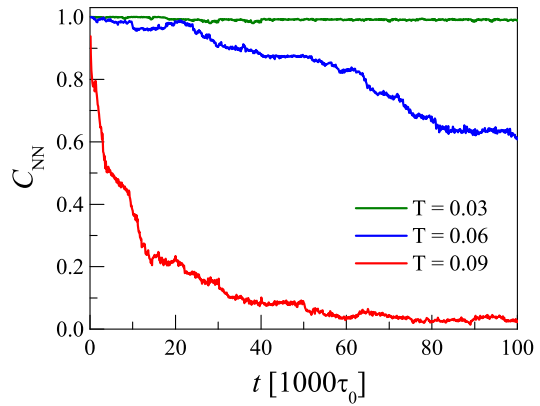}
\caption{Temporal evolution of number of nearest neighbors $C_{\rm
NN}$ kept from the initial state at $t=0$ (see the definition in
text) for $\alpha = 0.05$ at three typical temperatures. }
\end{center}
\end{figure}

We have also studied how vortices change their nearest neighbors
(NNs). At $t = 0$, we identify the NNs of each vortex in terms of
the Delaunay triangulation method. Due to random motions, some
vortices defined as NNs at $t=0$ diffuse away. The quantity $C_{\rm
NN}(t)$ is defined as the number of NNs identified at $t=0$
remaining as NNs up to time $t$, with a trivial normalization to
unity at $t=0$. The quantity $C_{\rm NN}$ varies with time in
qualitatively different ways in the three temperature regimes as
presented in Fig.~5: At $T=0.03$, $C_{\rm NN}$ remains unity as time
evolves where vortices are trapped in the initial cages and thus
NN's remain the same. At $T= 0.09$, $C_{\rm NN}$ decreases to zero
exponentially fast because the system is totally random. At
$T=0.06$, $C_{\rm NN}$ decreases slowly with time. Checking the
trajectories of vortices, we find that, in the intermediate
temperature regime $T_g<T<T_m$, diffusions of vortices take place in
terms of swapping nearest neighbors. Due to the mobility of vortices
at $T_g<T<T_m$ at equilibrium, the $v-F$ characteristics should be
linear even at infinitesimal driving force as revealed in Fig.~3.

Along with the random motions and neighbor swaps of vortices, we
also monitor the variation of crystalline order of the whole vortex
system. In Fig.~6, the structure factor $S(Q)$ at the first
reciprocal lattice vector of the triangular lattice and the fraction
of sixfold-coordinated vortices $P_6$ are displayed for the same
time duration of Fig.~5.  For $T=0.06$, the value of $S(Q)$ remains
constantly $\sim 0.3$ and Bragg peaks can be clearly observed (see
Fig.~4(b)). Therefore, the system always keeps the ordered
triangular structure with most vortices sixfoldly coordinated, even
though the individual vortices are mobile in a stochastic way.
Compared with thermal fluctuations, vortex-vortex interactions are
strong while pinning is weak, resulting in an unpinned BrG. The
pinned BrG at $T=0.03$ is well ordered as captured by large values
of $S(Q)$ and $P_6$, while they are both small at $T=0.09$
corresponding to the vortex liquid.

\begin{figure}[t]
\begin{center}
\includegraphics[width=8.5cm]{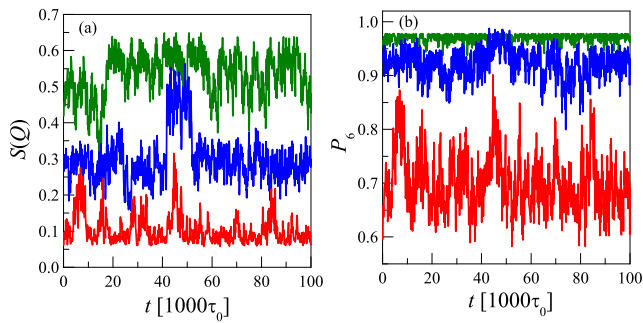}
\vspace{-3mm}\caption{Fluctuations of the structure factors at the
first reciprocal lattice vector $S(Q)$ (a) and fraction of
sixfold-coordinated vortices $P_6$ (b) for $\alpha = 0.05$ at three
typical temperatures. Green, blue and red are for $T=0.03$, 0.06,
and 0.09, respectively.}
\end{center}
\end{figure}

Recently an experimental finding of a second-order phase boundary
within the solid phase of the vortex system in BSCCO was reported
\cite{Beidenkopf}; the transition line $B_g(T)$ dividing the BrG
phase domain into two parts.  This transition was discussed as a
transition between the two different types of glasses, the
high-temperature \textit{pre-glass} or marginal glass phase, with
the nearly linear response behavior, and the low-temperature true
glass domain.  The transition between the two phases manifests
itself in a replica-symmetry breaking within the replica description
of disordered vortex system \cite{JSuper2006_LRV}. It is tempting to
associate our observation of the change in dynamic vortex behavior
with this transition. Indeed, at low temperatures $T < T_g$,
vortices remain indefinitely near their equilibrium positions since
on top of the elastic forces keeping them there, vortices are
additionally immobilized by disorder.  At intermediate temperatures,
$T_g < T < T_m$, pinning of single vortices is not efficient any
more and they may switch between their equilibrium positions due to
thermal diffusion.  However, putting this conclusion on a firm
quantitative basis requires a more detailed study of the processes
of vortex relaxation in both phases and will be a subject of a
forthcoming publication.

\vspace{3mm}

 \noindent {\it Acknowledgements --} The authors thank
M.~Tachiki, T.~Nattermann, E.~H.~Brandt, T.~Giamarchi,
B.~Rosenstein, H.~Beidenkopf, and A.~Tanaka for useful discussions.
Simulations were performed on HITACHI SR1100 at NIMS. This work has
been supported by WPI Initiative on Materials Nanoarchitectonics,
MEXT of Japan, CREST-JST of Japan, U.S. DoE Office of Science
(DE-AC02-06CH11357), and partially by ITSNEM of CAS.

\vspace{3mm} \noindent $\ast$ Corresponding author:
Hu.Xiao@nims.go.jp


\begin{thebibliography}{99}

\bibitem{Blatter_RMP} G. Blatter \textit{et al.}, Rev. Mod. Phys. \textbf{66}, 1125
(1994).

\bibitem{Ioffe1987} L. B. Ioffe and V. M. Vinokur, J. Phys. C \textbf{20}, 6149 (1987).

\bibitem{Vinokur2000} J. Kierfeld, H. Nordborg, and V. M. Vinokur, Phys. Rev. Lett.
\textbf{85}, 4948 (2000).

\bibitem{FRG} P. Chauve, T. Giamarchi, and P. Le Doussal, Phys. Rev. B
\textbf{62}, 6241 (2000).

\bibitem{Review_Nattermann2000} T. Nattermann and S. Scheidl, Adv. Phys. \textbf{49}, 607
(2000).

\bibitem{Review_Giamarchi} T. Giamarchi and S. Bhattacharya, \textit{ High Magnetic Fields:
Applications in Condensed Matter Physics, Spectroscopy} (Springer,
New York, 2002), p.314.

\bibitem{Review_Nattermann2004}  S. Brazovskii and T. Nattermann, Adv. Phys. \textbf{53}, 177 (2004).

\bibitem{LO} A. I. Larkin and Yu. N. Ovchinnikov, J. Low Temp. Phys. \textbf{34},
409 (1979).

\bibitem{Nattermann1987} T. Nattermann, Europhys. Lett. \textbf{4}, 1241 (1987).

\bibitem{Fisher1989} M. P. A. Fisher, Phys. Rev. Lett. \textbf{62}, 1415 (1989).

\bibitem{Feigelman1989} M. V. Feigel'man \textit{et al.}, Phys. Rev. Lett.
\textbf{63}, 2303 (1989).

\bibitem{BrG_Nattermann} T. Nattermann, Phys. Rev. Lett. \textbf{64}, 2454 (1990).

\bibitem{Lemerle_1998} S. Lemerle \textit{et al.}, Phys. Rev. Lett.  \textbf{80}, 849
(1998).

\bibitem{Kolton} A. B. Kolton, A. Rosso, and T. Giamarchi, Phys. Rev. Lett.
\textbf{94}, 047002 (2005).

\bibitem{BrG_GD} T. Giamarchi and P. Le Doussal, Phys. Rev. Lett. \textbf{72}, 1530
(1994); Phys. Rev. B \textbf{52}, 1242 (1995).

\bibitem{VG} D. S. Fisher, M. P. A. Fisher, and D. A. Huse, Phys. Rev. B \textbf{43}, 130 (1991).

\bibitem{model} E.~H.~Brandt, Phys. Rev. Lett. \textbf{50}, 1599 (1983);
{\it ibid} J. Low Temp. Phys. \textbf{53}, 41 (1983); S.~Ryu
\textit{et al.}, Phys. Rev. Lett. \textbf{68}, 710 (1992);
C.~Reichhardt, C.~J.~Olson, and F. Nori, Phys. Rev. Lett.
\textbf{78}, 2648 (1997); A.~van~Otterlo, R.~T.~Scalettar, and
G.~T.~Zim\'anyi, Phys. Rev. Lett. \textbf{81}, 1497 (1998); E.~Olive
\textit{et al.}, Phys. Rev. Lett. \textbf{91}, 037005 (2003).

\bibitem{PRL2007_LH} M. B. Luo and X. Hu, Phys. Rev. Lett. \textbf{98}, 267002 (2007).

\bibitem{Bulaevskii_2004} L. N. Bulaevskii \textit{et al.}, Phys. Rev. B
\textbf{50}, 3507 (1994).

\bibitem{Blatter2004} G.~Blatter, V.~B.~Geshkenbein, and
J.~A.~G.~Koopmann, Phys. Rev. Lett. \textbf{92}, 067009 (2004).

\bibitem{PRB1997_GD} T. Giamarchi and P. Le Doussal, Phys. Rev. B
\textbf{55}, 6577 (1997).

\bibitem{Beidenkopf} H. Beidenkopf \textit{et al.}, Phys. Rev. Lett. \textbf{95},
257004 (2005).

\bibitem{JSuper2006_LRV} D.~Li, B.~Rosenstein, and V.~Vinokur, J.
Supercond. Nov. Magn. \textbf{19}, 369 (2006).

\end{thebibliography}
\end{document}